\documentclass[a4paper,11pt]{article}
\pdfoutput=1

\usepackage{jheppub}

\usepackage{hyperref}
\usepackage{graphicx}
\usepackage{amsmath}
\usepackage{amssymb}
\usepackage{xspace}
\usepackage{mathrsfs}
\usepackage{subfigure}
\usepackage{slashed}
\usepackage[normalem]{ulem}
\usepackage{xcolor} 
\usepackage{epstopdf}
\usepackage{soul}

\DeclareRobustCommand{\eq}[1]{eq.~\eqref{eq:#1}}

\DeclareRobustCommand{\eqss}[2]{eqs.~\eqref{eq:#1} -- \eqref{eq:#2}}
\DeclareRobustCommand{\fig}[1]{fig.~\ref{fig:#1}}

\DeclareRobustCommand{\sec}[1]{sec.~\ref{sec:#1}}

\DeclareRobustCommand{\rcite}[1]{ref.~\cite{#1}}
\DeclareRobustCommand{\rcites}[1]{refs.~\cite{#1}}

\usepackage{marginnote}

\newcommand{\ord}[1]{\mathcal{O}\!\left(#1\right)}

\newcommand{\df}{\mathrm{d}}

\DeclareMathOperator{\tr}{tr}

\newcommand{\eps}{\epsilon}

\newcommand{\nn}{\nonumber}

\newcommand{\cusp}{\mathrm{cusp}}

\newcommand\lr[1]{{\left({#1}\right)}}
\newcommand\Cfour{\mathcal{C}_{F,4}}

\def\nn{\nonumber}
\def\df{\textrm{d}}

\def\MS{\overline{\rm MS}}

\allowdisplaybreaks[2]

\preprint{\begin{flushright}
MITP/17-050
\end{flushright}}

\title{
On the Casimir scaling violation in the cusp anomalous dimension at small angle.
}

\author[a,b,c]{Andrey Grozin,}
\author[c]{Johannes Henn,}
\author[c]{Maximilian Stahlhofen}

\affiliation[a]{Budker Institute of Nuclear Physics SB RAS, Novosibirsk 630090, Russia}
\affiliation[b]{Novosibirsk State University, Novosibirsk 630090, Russia}
\affiliation[c]{PRISMA Cluster of Excellence, Johannes 
Gutenberg University, 55128 Mainz, Germany}

\emailAdd{A.G.Grozin@inp.nsk.su}
\emailAdd{henn@uni-mainz.de}
\emailAdd{mastahlh@uni-mainz.de}

\abstract{
We compute the four-loop $n_f$ contribution proportional to the quartic Casimir 
of the QCD cusp anomalous dimension as an expansion for small cusp angle 
$\phi$. This piece is gauge invariant, violates Casimir scaling, and first 
appears at four loops.
It requires the evaluation of genuine non-planar four-loop Feynman integrals.
We present results up to ${\mathcal O}(\phi^4)$.
One motivation for our calculation is to probe a recent conjecture on the 
all-order structure of the cusp anomalous dimension. 
As a byproduct we obtain the four-loop HQET wave function anomalous dimension for this color structure.
}

\begin{document}
\maketitle

\section{Introduction}
\label{sec:Intro}

The cusp anomalous dimension $\Gamma_\cusp(\phi)$, defined as the anomalous dimension of a Wilson loop with a cusp of angle $\phi$ \cite{Polyakov:1980ca}, determines the renormalization group evolution of the Isgur--Wise function~\cite{Falk:1990yz}.
In this paper we will mostly be interested in the small $\phi$ expansion of $\Gamma_\cusp(\phi)$.
Such an expansion is performed for extracting the CKM matrix element $V_{cb}$ from the $B \to D^*$ semileptonic decays, see e.g. \rcite{Amhis:2016xyh}:
The extrapolation of experimental points to $\phi=0$ is done using the slope and curvature of the Isgur--Wise function, i.e. its $\phi^2$ and $\phi^4$ terms.
The cusp anomalous dimension at small angle is also related to real gluon radiation in the case when a heavy quark slightly changes its velocity.
This kind of radiation has been considered in~\rcites{Czarnecki:1997sz,Correa:2012at}.

The QCD cusp anomalous dimension is currently known to three
loops~\cite{Polyakov:1980ca,Korchemsky:1987wg,Correa:2012nk,Grozin:2014hna,Grozin:2015kna} for arbitrary angle $\phi$.
Up to this order, the anomalous dimensions for Wilson lines in a given representation $R$ of the color group are related by Casimir scaling: They are given by the quadratic Casimir operator $C_R$ times a universal ($R$-independent) function.
Here we explicitly demonstrate that this is not so at four loops.
We consider a cusp with a small angle $\phi$,
and calculate the first terms of the expansion of its anomalous dimension $\Gamma_\cusp(\phi)$,
namely the $\phi^2$ and $\phi^4$ terms.
We consider the specific color structure $n_f \Cfour$ with
\begin{equation}
\Cfour \equiv \frac{d_R^{abcd}d_F^{abcd}}{N_R}\,,
\quad\text{where}\quad
d_F^{abcd} = \tr_F\bigl[T_F^{(a} T_F^{b\vphantom{(}} T_F^{c\vphantom{(}} T_F^{d)}\bigr]\,,\quad
d_R^{abcd} = \tr_R\bigl[T_R^{(a} T_R^{b\vphantom{(}} T_R^{c\vphantom{(}} T_R^{d)}\bigr]\,.
\label{eq:C4}
\end{equation}
The $T_F^a$ denote the gauge group generators in the fundamental representation,
the $T_R^a$ the ones in the representation $R$ of dimensionality $N_R = \tr_R \mathbf{1}$,
and the round brackets indicate symmetrization, see~\rcite{vanRitbergen:1998pn}.
This color structure cannot be represented as $C_R$ times a universal constant,
and thus breaks Casimir scaling.

Non-zero quartic Casimir contributions are known to occur in closely related quantities, such as the static quark anti-quark potential~\cite{Anzai:2010td,Lee:2016cgz}, which corresponds to the $\phi \to \pi$ limit of $\Gamma_\cusp(\phi)$.
Also in $\mathcal{N}=4$ super Yang--Mills (sYM) theory contributions proportional to the quartic Casimir were found in the Bremsstrahlung function~\cite{Correa:2012at}, i.e. the $\phi^2$ term, and, very recently, in the light-like limit~\cite{Boels:2017skl} of $\Gamma_\cusp$.

Up to three loops the cusp anomalous dimension has an interesting property~\cite{Grozin:2014hna,Grozin:2015kna}:
When expressed in terms of an effective coupling constant $a$, which is defined such that the large Minkowskian $\phi$ asymptotics, i.e. the light-like limit, of $\Gamma_\cusp(\phi)$ is given by the first-order $a$ term only, it becomes a universal function $\Omega(\phi,a)$ that is independent of the number of fermion or scalar fields in the theory.
It has been conjectured in \rcites{Grozin:2014hna,Grozin:2015kna} that this property holds to all orders of perturbation theory, simply from the intriguing empirical observation at the first three orders.
In the present paper we check this conjecture at four loops.
The $n_f \Cfour$ term we are interested in contains the number of massless fermions $n_f$, and, according to the conjecture, can only arise
from some $\alpha_s^n$ ($n>1$) term in $a$.
It cannot be represented as a product of lower-loop color structures,
and hence it can only come from the term $c\, n_f \Cfour/C_R (\alpha_s/\pi)^4$ in $a/\pi$
inserted in the leading term $\Gamma_\cusp(\phi) = C_R (a/\pi) (\phi \cot\phi - 1) + \ord{a^2}$.
The normalization factor $c$ can be determined from the limit $\phi\to\pi$,
where the four-loop $\Gamma_\cusp$ is related~\cite{Grozin:2014hna,Grozin:2015kna}
to the three-loop static potential~\cite{Anzai:2009tm,Lee:2016cgz,Smirnov:2009fh}.

We find that the analytic form of our result is different from the 
conjecture of \rcites{Grozin:2014hna,Grozin:2015kna}.
Interestingly, the numerical values are still surprisingly close to the conjectured ones.
While this paper was finalized, 
the light-like QCD cusp anomalous dimension at four loops
has been computed numerically~\cite{Moch:2017uml}. 
Its $n_f \Cfour$ term is also relatively close, but different from the conjecture. 
This is in line with our findings here.

As a by-product of our calculation (at $\phi =0$), 
we determine  the $n_f \Cfour$ term in the four-loop anomalous dimension 
of the HQET heavy-quark field. Currently it is only known to three loops~\cite{Melnikov:2000zc,Chetyrkin:2003vi}.
Our result can serve as a non-trivial cross-check of future calculations.

The paper is organized as follows. In \sec{calculation}, we describe our calculation. In \sec{HQET} we compute the
heavy quark anomalous dimension and extract from it the QCD on-shell heavy-quark field renormalization constant. In \sec{results} we present the results for the cusp anomalous dimension to order $\phi^4$, and compare to the conjecture of \rcites{Grozin:2014hna,Grozin:2015kna}.

\section{Calculation}
\label{sec:calculation}

The QCD cusp anomalous dimension arises from the UV divergences of the Wilson 
loop
\begin{align}
W = \frac{1}{N_R} \big\langle0|\tr_R\, P\, \exp \lr{i g\oint_C dx^\mu 
A_\mu(x)}|0\big\rangle = 1+ \ord{g^2}\,,
\label{def}
\end{align}
where $A_\mu=A_\mu^a\,T_R^a$ is the gluon field, $P$ is the path-ordering 
operator, the trace is over (color) indices in the representation $R$ of the 
gauge group.
The closed integration contour $C$ has a cusp at a single point and is smooth 
otherwise.
Without loss of generality, we can choose the contour $C$ to consist of two 
Wilson lines along the directions $v_1^\mu$ and $v_2^\mu$ with $v_1^2=v_2^2=1$ 
that both extend to infinity and end at the cusp point.
We denote the angle between them by $\phi$, where $\phi=0$ corresponds to a 
Wilson line along $v_1^\mu = v_2^\mu$ with both ends at infinity, and 
\begin{align}
\cos \phi = v_1 \cdot v_2 \,.
\end{align}
The open ends of the Wilson lines are considered to be closed at infinity. They 
can be interpreted as heavy quark lines in HQET with $v_1^\mu$ and $v_2^\mu$ 
being the heavy quark velocities.
We note that for real $v_i$ in Minkowski spacetime, $\phi=i \varphi$ is purely 
imaginary and $\cosh \varphi = v_1 \cdot v_2$.
In this configuration the cusp anomalous dimension was computed through three 
loops in \rcites{Polyakov:1980ca,Korchemsky:1987wg,Correa:2012nk,Grozin:2014hna,Grozin:2015kna} and we refer to the latter 
reference for details on the calculational setup.\footnote{Partial results at four loops in $\mathcal{N}=4$ super Yang--Mills are also available, see \rcites{Correa:2012at,Henn:2013wfa}.}

We distinguish two types of HQET Feynman diagrams contributing to the 
Wilson loop $W$ beyond tree-level: heavy quark self-energy and 
one-particle-irreducible (cusp) vertex correction diagrams. 
The sum of the latter depends on the angle $\phi$ and is denoted by $V(\phi)$.
Via a simple Ward identity the self-energy can be related to the vertex 
correction at $\phi=0$.
We can thus write~\cite{Korchemsky:1987wg}
\begin{align}\label{eq:lnW}
\log W = \log V(\phi) - \log V(0) = \log Z + \ord{\eps^0}\,,
\end{align}
where we have introduced the cusp renormalization factor $Z$. Here and 
throughout this paper we use dimensional regularization with $d=4-2\eps$.
The cusp anomalous dimension is then given by
\begin{align}\label{eq:cusp-def}
\Gamma_{\rm cusp}(\phi,\alpha_s) 
= \frac{\df\log Z}{\df\log\mu} 
\,.
\end{align}

\begin{figure*}[t]
\begin{center}
\includegraphics[width=4.5 cm]{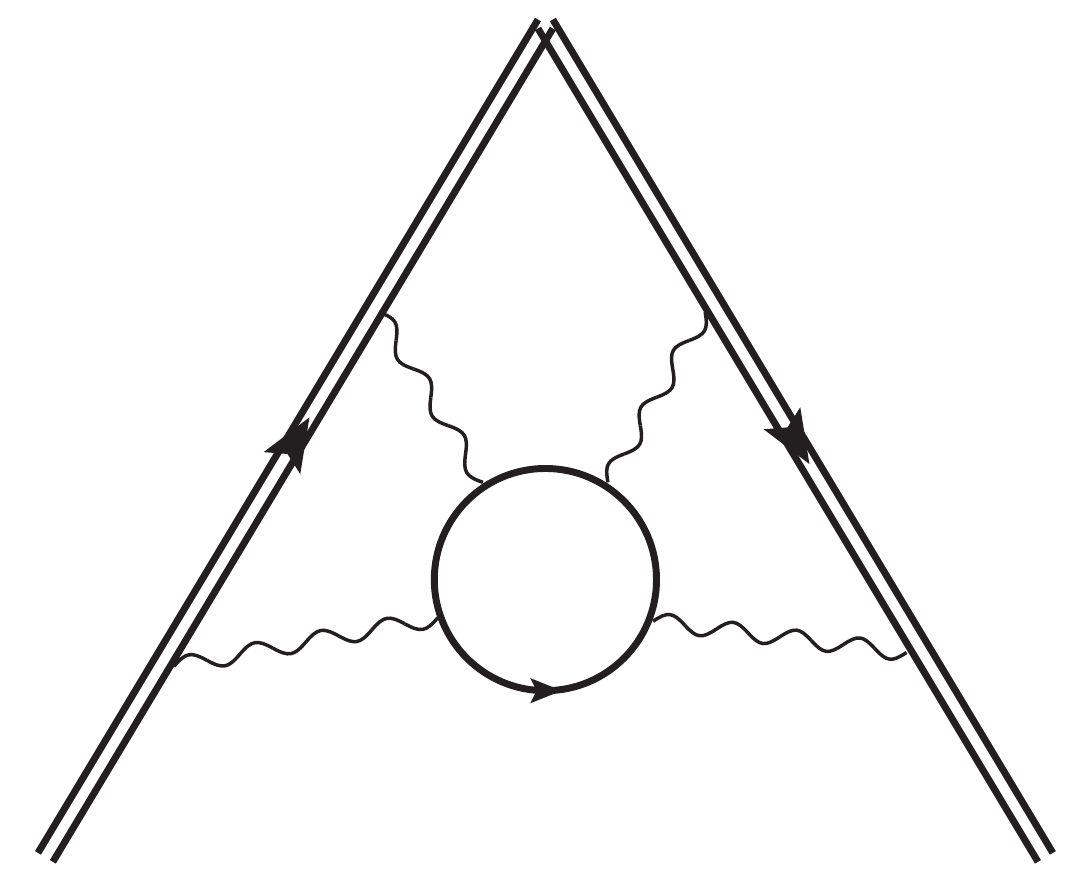}%
\put(-130,20){a)}
\put(-110,55){$v_1$}
\put(-28,55){$v_2$}
\qquad
\includegraphics[width=4.5 cm]{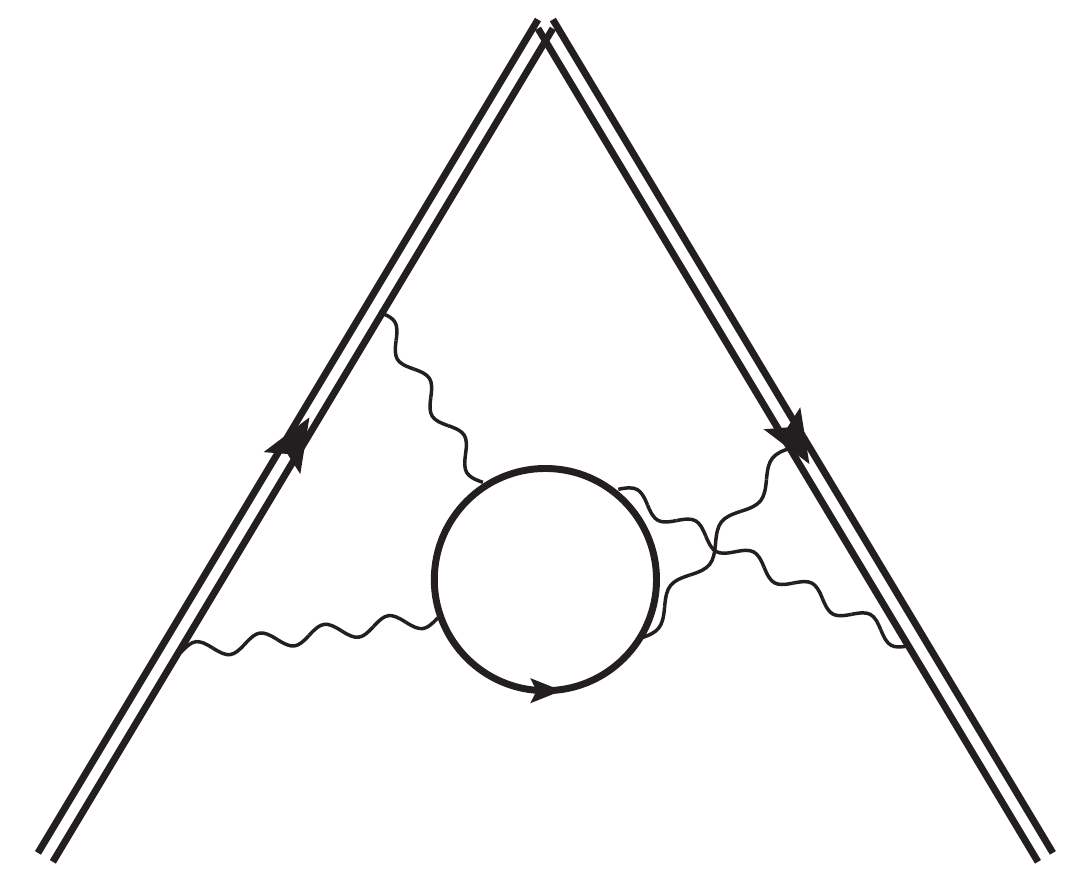}%
\put(-130,20){b)}
\qquad
\includegraphics[width=4.5 cm]{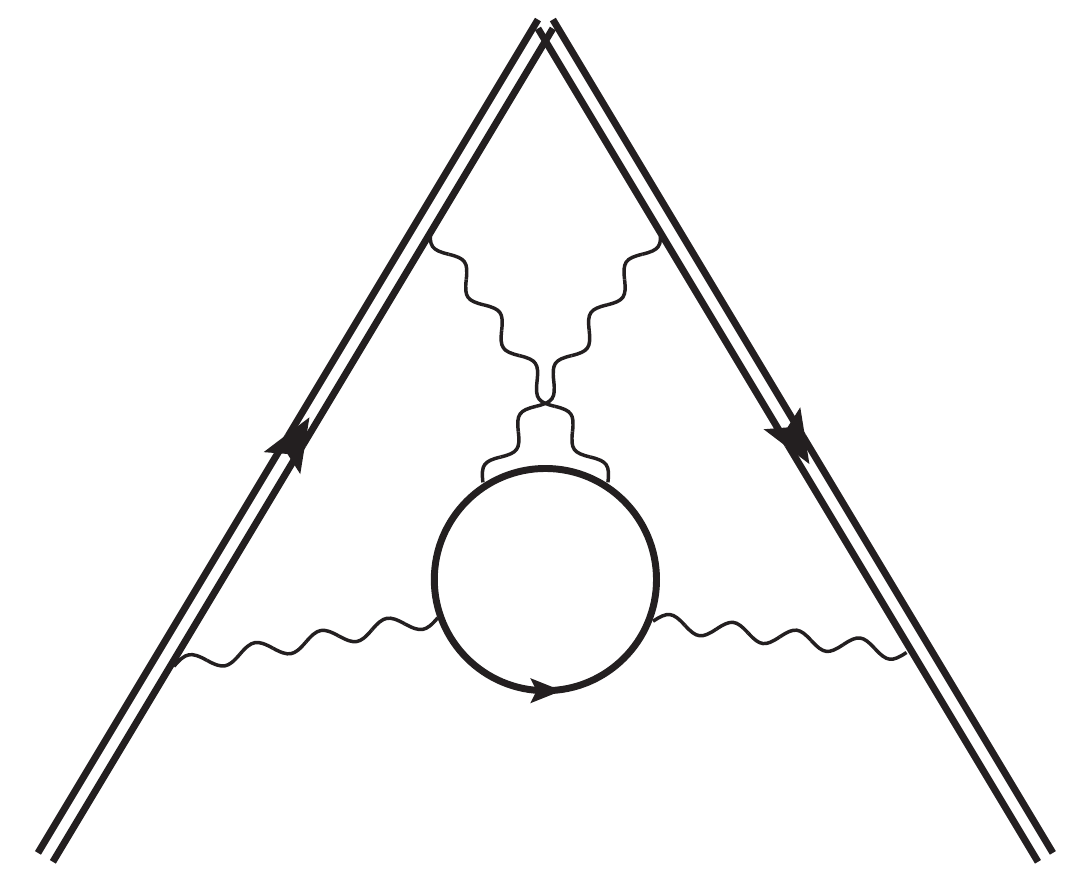}
\put(-130,20){c)}
\\
\includegraphics[width=4.5 cm]{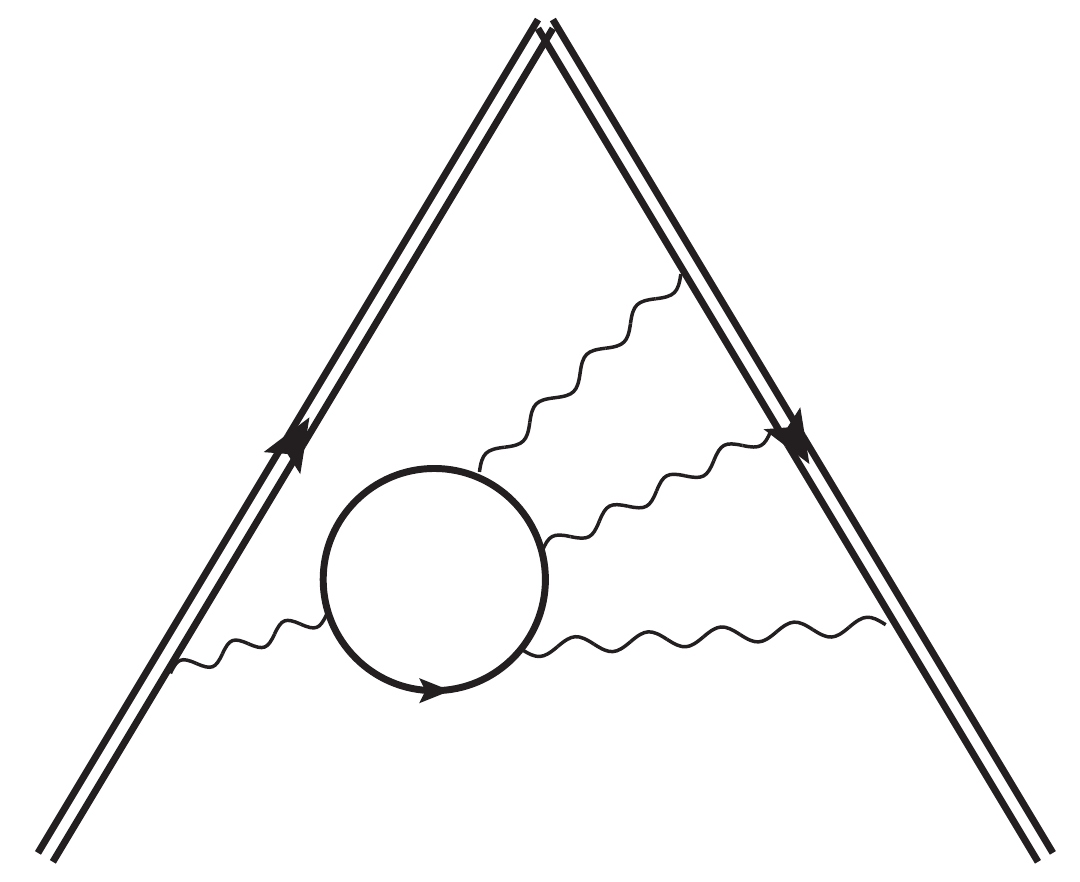}%
\put(-130,20){d)}
\qquad
\includegraphics[width=4.5 cm]{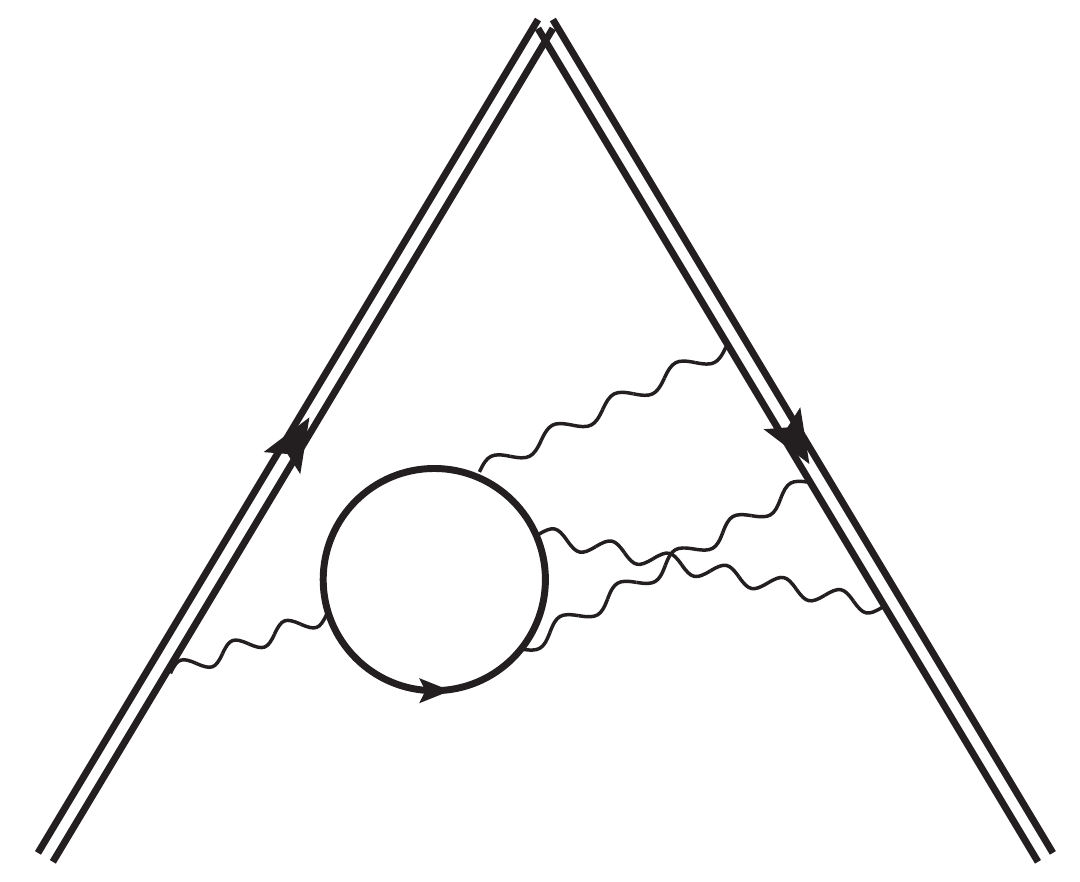}%
\put(-130,20){e)}
\qquad
\includegraphics[width=4.5 cm]{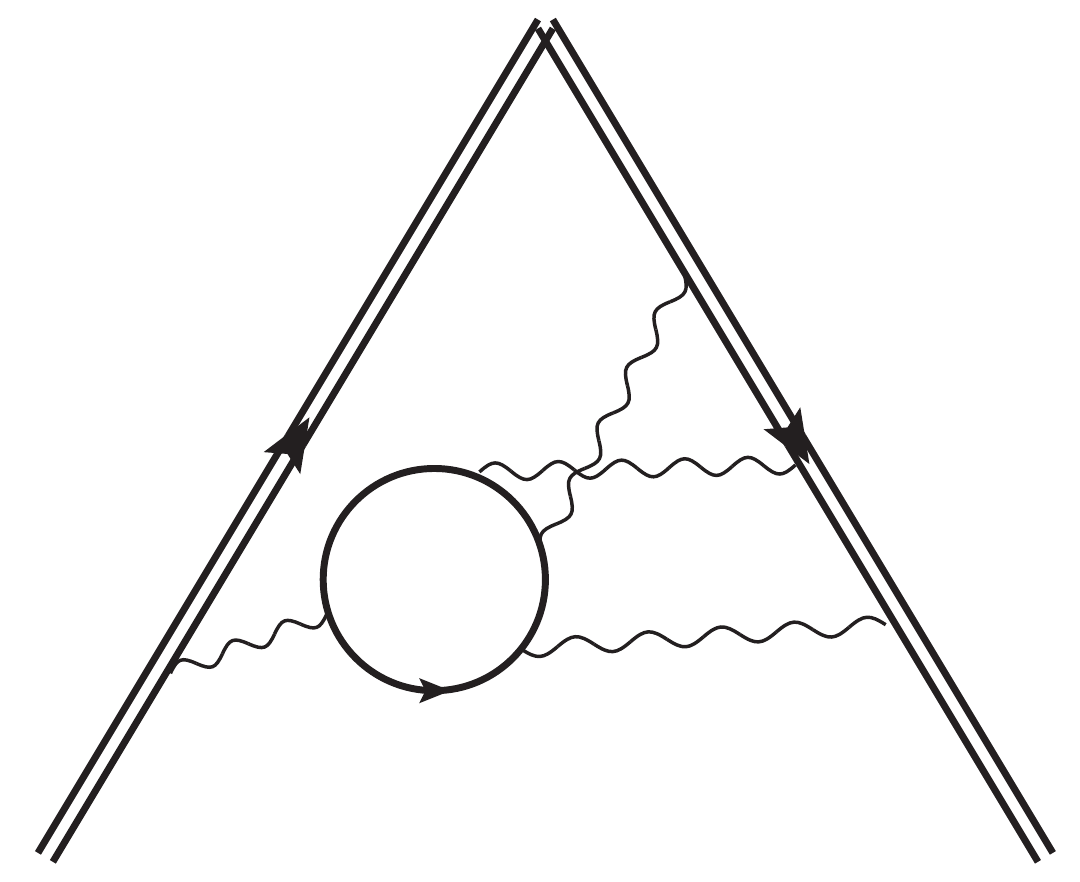}%
\put(-130,20){f)}
\end{center}
\caption{Diagrams contributing to the $n_f \Cfour$ term in the  vertex function 
$V(\phi)$. Double lines represent Wilson lines, wavy lines gluons and single 
lines the $n_f$ light quarks.  
Left-right mirror graphs and the diagrams with reversed light fermion 
flow are not displayed.
The $n_f \Cfour$ contributions of the latter equal the ones of 
their relatives shown here.
\label{fig:diags}
}
\end{figure*}

We are interested in color structures that generate the quartic Casimir $\Cfour$ defined in \eq{C4}.
It first appears in the QCD vertex correction $V$ at four loops and violates 
Casimir scaling.
We will use the equality~\cite{vanRitbergen:1998pn}
\begin{align}
\Cfour = \frac{1}{N_R} \, \tr_R \big[T_R^a T_R^b T_R^c T_R^d \big] \;
\tr_F \big[T_F^a T_F^b T_F^c T_F^d \big] + \ldots\,,
\label{eq:C4eq}
\end{align}
where the ellipsis in \eq{C4eq} stands for terms that can be expressed only in 
terms of the quadratic Casimirs $C_R$, $C_F$ and $C_A$.
Equation~\eqref{eq:C4eq} also holds when the order of the adjoint color indices 
($a,b,c,d$) in one of the traces on the right-hand side is interchanged 
arbitrarily.

The quartic Casimir $\Cfour$ occurs in the four-loop contribution proportional 
to $n_f$. The latter denotes the number of light (massless) fermions in the fundamental 
representation $F$.
From \eq{C4eq} it is clear that the four-loop Feynman diagrams involving 
$\Cfour$ must have a light fermion loop forming a box that is connected to the Wilson lines via four gluons.
There are only six different diagram topologies of that type contributing to 
$V$.
They are displayed in \fig{diags}.
Counting also diagrams with reversed light fermion flow and left-right mirror 
graphs we arrive at a total of $18$ 
diagrams that contribute to the $\Cfour$ 
term.%
\footnote{We note that there is one more Casimir scaling violating color structure at 
four loops, namely $d_R^{abcd}d_A^{abcd}/N_R$. It arises in the purely 
gluonic correction to $V$ ~\cite{Grozin:2015kna,Anzai:2010td}. The 
number of involved diagram topologies is however much bigger than in the case of 
$\Cfour$.}
Up to the three-loop order an analysis of all color structures and taking into
account non-Abelian exponentiation~\cite{Gatheral:1983cz,Frenkel:1984pz} 
makes it possible to rewrite all non-planar integrals in terms of planar integrals only~\cite{Grozin:2015kna}. 
This is not the case for the non-planar diagrams ($b$\,-$f$).

Using the HQET building blocks ($i=1,2$)
\begin{align}
S(k) = \frac{\rlap/k}{k^2}\,,\quad
H_i(k) = \frac{1}{k\cdot v_i - \frac{\delta}{2}}\,,\quad
V_i(k) = \frac{1}{k^2} \bigg[ \rlap/v_i - \xi \frac{k\cdot v_i}{k^2} \rlap/k 
\bigg] ,
\end{align}
associated with the fermion, heavy quark and gluon lines, 
respectively, we can write the diagrams of \fig{diags} in generalized covariant 
gauge ($\xi=0$ corresponds to Feynman gauge) in compact form.
The off-shellness $\delta/2$ in the heavy quark propagators serves as an 
infrared regulator~\cite{Grozin:2015kna} and can be interpreted as an 
external energy flowing in the opposite direction of the $v_i^\mu$ (indicated 
by the arrows in \fig{diags}).
For the coefficients of the color factor $n_f \Cfour$ we have
\begin{align}
D_a &= - g^8 H_1(k_2) H_1(k_3) H_2(k_2)H_2(k_1) \,\tr\big[
S(k_4)V_1(k_3)S(k_4-k_3)V_1(k_2-k_3) S(k_4-k_2)
\nn\\
&\quad \times V_2(k_1-k_2) S(k_4-k_1)V_2(-k_1)\big]
\,, \label{eq:diaga}\\
D_b &= - g^8 H_1(k_2) H_1(k_3) H_1(k_1)H_2(k_1) \, \tr\big[
S(k_4)V_2(k_1)S(k_4-k_1)V_1(k_2-k_1)S(k_4-k_2)
\nn\\
&\quad \times V_1(k_3-k_2)S(k_4-k_3)V_1(-k_3)\big]
\,,\\
D_c &= - g^8 H_1(k_2) H_1(k_3) H_2(k_2) H_2(k_1) \, \tr\big[
S(k_4)V_1(k_2-k_3)S(k_4-k_2+k_3)V_1(k_3)S(k_4-k_2)
\nn\\
&\quad \times V_2(k_1-k_2)S(k_4-k_1)V_2(-k_1) \big]
\,,\\
D_d &= - g^8 H_1(k_2) H_1(k_3) H_1(k_1) H_2(k_1) \, \tr\big[
S(k_2-k_4)V_1(k_2-k_1)S(k_1-k_4)V_2(k_1)S(-k_4)
\nn\\
&\quad \times V_1(k_3-k_2)S(k_2-k_3-k_4) V_1(-k_3) \big]
\,,\\
D_e &= - g^8 H_1(k_2) H_1(k_3) H_2(k_2) H_2(k_1) \, \tr\big[
S(k_4)V_1(k_3)S(k_4-k_3)V_2(k_1-k_2)S(k_2-k_3+k_4-k_1)
\nn\\
&\quad \times V_1(k_2-k_3)S(k_4-k_1)V_2(-k_1) \big]
\,,\\
D_f &= - g^8 H_1(k_2) H_1(k_3) H_1(k_1) H_2(k_1) \, \tr\big[
S(k_4)V_2(k_1)S(k_4-k_1)V_1(k_3-k_2)S(k_2-k_3+k_4-k_1)
\nn\\
&\quad \times V_1(k_2-k_1)S(k_4-k_3) V_1 ( -k_3)\big]
\,. \label{eq:diagf}
\end{align}
The overall minus sign in the above expressions originates from the closed 
fermion loop.
The sum of all 18 contributions to the $n_f \Cfour$ term is gauge invariant and 
reads%
\footnote{Although maybe not immediately obvious in \fig{diags}, diagram $b$ is just 
as symmetric as diagrams $a$ and $c$, once both light fermion flows are taken 
into account. This can e.g. be seen by flipping or twisting the light quark 
loop. Thus there is no factor of two in front of $D_b$ from a left-right mirror 
graph in \eq{Vdiag}.}
\begin{align}
V(\phi)\big|_{n_f \Cfour} &= 2\, n_f \Cfour \,  \big(    
D_a+ D_b+ D_c + 2D_d + 2D_e+ 2D_f
\big)\,.
\label{eq:Vdiag}
\end{align}
The gauge invariance can be seen as follows:
The $n_f \Cfour$ contribution in \eq{Vdiag} is effectively QED-like, as all 
diagrams have the same color structure (no relative factors).
Now, consider the subdiagrams consisting of the fermion loop and four off-shell 
gluons attached to it in all possible ways. If we pick out one of the gluon 
vertices and contract it with the four-momentum of the associated gluon the 
contributions from the 18 one-loop diagrams add up to zero owing to the 
Ward-Takahashi identity of QED, see e.g. \rcite{Peskin:1995ev}.
Despite being off-shell the gluons are therefore effectively transverse. Thus 
the (longitudinal) $\xi$ terms in their propagators vanish in the sum of all 18 
diagrams.
Moreover, renormalization group consistency requires \eq{Vdiag} to be at most 
$1/\eps$ divergent (for finite $\phi$), as there are no (UV) divergent 
subdiagrams involved.
We will use these properties as strong cross-checks of our calculation.

In this paper we are interested in the expansion of the cusp anomalous 
dimension for small angle $\phi$.
The calculation of the individual terms in this $\phi$ expansion is technically 
simpler than the calculation for arbitrary $\phi$, as there is only one 
external vector $v^\mu=v_1^\mu=v_2^\mu$ and the results are pure numbers. 
It can therefore be considered as a first step toward the calculation of the full 
angle-dependent cusp anomalous dimension, but also directly yields some relevant 
physical information as outlined in \sec{Intro}.

Unlike e.g. the parallel lines limit ($\phi \to \pi$) or the light-like limit 
($\phi \to i \infty$) the small angle limit is 
well-behaved~\cite{Correa:2012nk,Grozin:2015kna}, i.e. we can safely expand in 
$\phi$ before integration over the loop momenta.
By virtue of \eq{lnW} the leading order (LO) term ($\propto \phi^0$) vanishes.

In practice the Taylor expansion of the expressions in \eqss{diaga}{diagf} can 
e.g. be done as follows: We write (in Euclidean spacetime) $v_1=v$, $v_2 = 
\cos(\phi) v + \sin(\phi) e_2$ with $v^2 = e_2^2 = 1$, $v\cdot e_2 = 0$ and 
differentiate the integrands w.r.t. $\phi$.
The numerators of the resulting terms include a number of scalar products of 
the loop momenta $k_i^\mu$ with the unit vector $e_2^\mu$. The denominators are 
free of such scalar products. Upon integration therefore terms with odd numbers 
of $k_i \cdot e_2$, i.e. odd powers of $\phi$, vanish because of the 
antisymmetry of the integrand. For the terms with even numbers of $k_i \cdot 
e_2$ we perform a tensor reduction. After evaluation of the Dirac trace we end up 
with a scalar integrand that only involves the 14 independent scalar products 
$k_i^2$, $k_i \cdot k_j$ and $k_i \cdot v$.
The result for each diagram can thus be expressed as a linear combination of 
integrals
\begin{align}
&G_{a_1,\ldots,a_{14}} = e^{4 \eps \gamma_E} \int \!\! \frac{\df^d k_1}{ 
i\pi^{d/2}} \int \!\! \frac{\df^d k_2}{ i\pi^{d/2}}
\int \!\! \frac{\df^d k_3}{ i\pi^{d/2}} \int \!\! \frac{\df^d k_4}{ i\pi^{d/2}} 
\;\prod_{i=1}^{14} (Q_i)^{-a_i}\,,
\label{eq:Gdef}
\end{align}
with
\begin{align}
&Q_1 =  - 2 k_1 \cdot v + \delta\,,\;
Q_2 =  - 2 k_2 \cdot v + \delta\,,\;
Q_3 =  - 2 k_3 \cdot v + \delta\,,\;
Q_4 =  - k_1^2\,,\;
Q_5 =  - (k_1-k_2)^2\,,\;
\nn\\
&Q_6= -(k_2 - k_3)^2 \,,\;
Q_7= -k_3^2 \,,\;
Q_8= -k_4^2 \,,\;
Q_9= -(k_1 - k_4)^2\,,\;
Q_{10}=-(k_2 - k_4)^2\,,\;
\nn\\
&Q_{11}=-(k_3 - k_4)^2\,,\;
Q_{12}=-(k_2 - k_3 - k_4)^2\,,\;
Q_{13}= -(k_1 - k_2 + k_3 - k_4)^2\,,\;
\nn\\
&Q_{14}= - 2 k_4 \cdot v + \delta\,,
\end{align}
at every order in $\phi$.
The indices $a_i$ are integer numbers, which for $1 \le i \le 13$ can be 
positive (propagators), zero, or negative (numerators), while $a_{14} \le 0$, 
i.e. $Q_{14}$ only appears as a numerator in our problem.
Note that for brevity we have suppressed the usual $-i0$ Feynman prescription 
in the $Q_i$, which is needed to ensure causality in Minkowski spacetime.

\begin{figure*}[t]
\begin{center}
\includegraphics[width=5.13 cm]{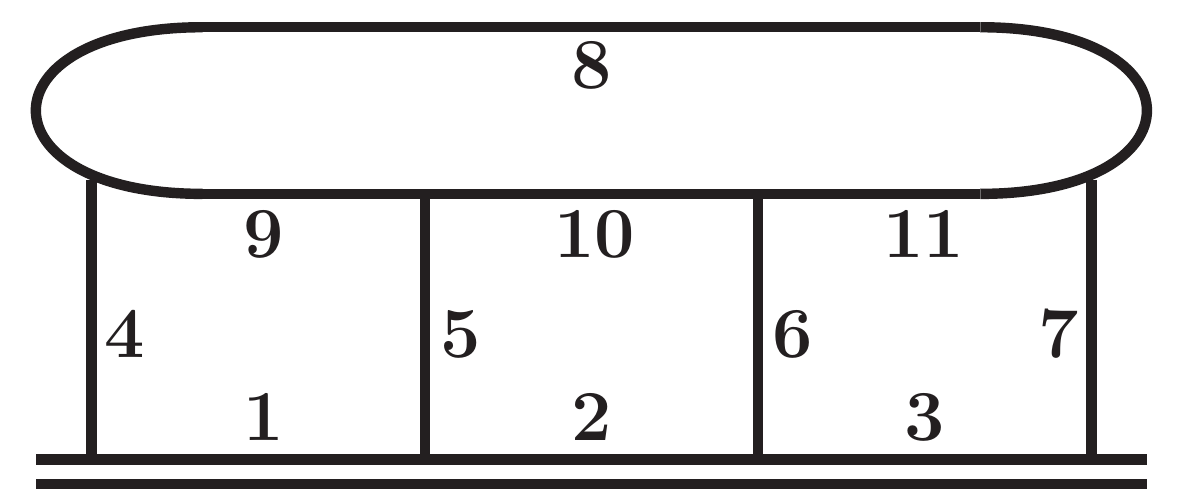}%
\put(-80,-12){(1)}
\,
\includegraphics[width=5.13 cm]{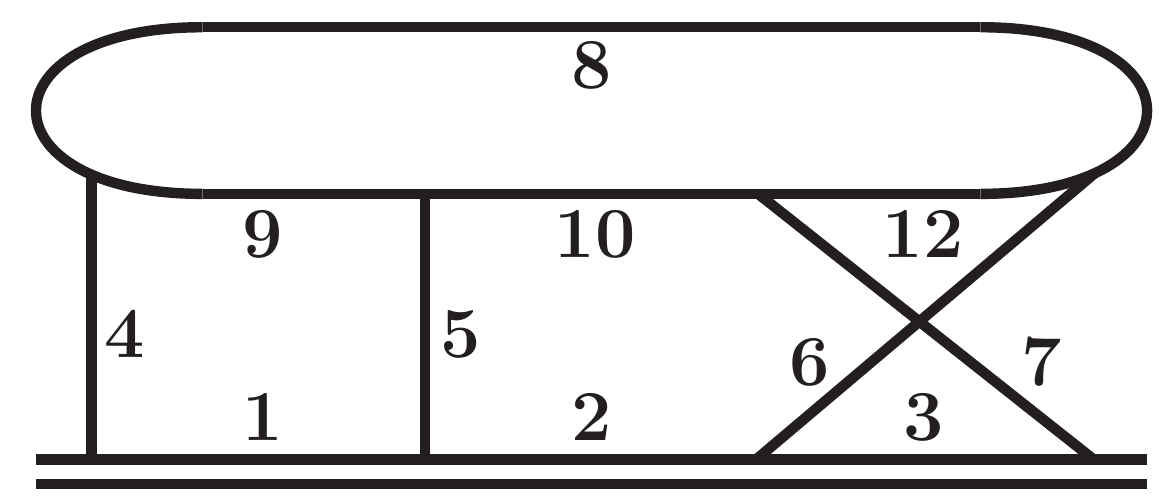}%
\put(-80,-12){(2)}
\,
\includegraphics[width=5.13 cm]{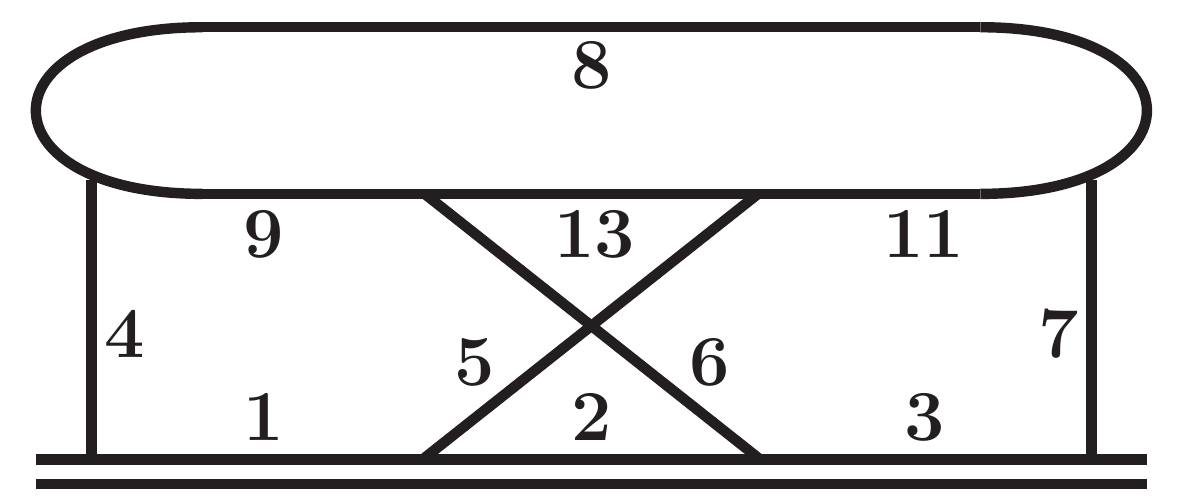}
\put(-80,-12){(3)}
\end{center}
\caption{
Independent integral topologies (1,2,3) for the small angle expansion of 
$V(\phi)$.
The number $i$ for each (double) line refers to the corresponding propagator 
with power $a_i$ in $G_{a_1,\ldots,a_{14}}$ according to \eq{Gdef}.
Propagators 1,2,3 are (linear) Wilson line (heavy quark) propagators.
\label{fig:topos}
}
\end{figure*}

The integrals $G$ contributing to the vertex function $V$ have at most 11
propagators. 
According to its propagator configuration each of them can be assigned to one 
of the integral topologies displayed in \fig{topos}.
The natural way to do this is to map the small angle expanded diagrams 
$D_{a,d}$ onto topology 1, $D_{b,e}$ onto topology 2, and $D_{c,f}$ onto 
topology 3.\footnote{For integrals with less than 11 propagators this assignment 
may not be unique.}
We can now perform an integration-by-parts (IBP) 
reduction~\cite{Chetyrkin:1981qh} for the integrals of each topology 
separately.
We do this reduction to master integrals (MI) for all relevant integrals 
up to $\ord{\phi^4}$.
To this end, we use the public computer program {\tt 
FIRE5}~\cite{Smirnov:2014hma} in combination with {\tt 
LiteRed}~\cite{Lee:2012cn,Lee:2013mka}.
In this way, we find 32 MI for topology 1, 32 MI for topology 2 and 
30 MI for topology 3.
Taking into account relations among the MI of different topologies (with less than 11 
propagators), the total number of independent MI across the three topologies is 
43.
Expressing the $\phi$-expanded vertex function $V$ as a linear 
combination of these 43 MI, we explicitly verify that the 
$\xi$-dependence in the coefficients of the MI drops out at $\ord{\phi^0}$ and 
$\ord{\phi^2}$ as required by gauge invariance.
At order $\ord{\phi^4}$, we restrict ourselves to the 
Feynman gauge for performance reasons.
We note that the extension to higher powers of $\phi$ 
is conceptionally straightforward, and is only limited by
computing resources.

As the final step of our calculation we have to solve the MI to 
sufficiently high order in $\eps$ in order to determine the overall $1/\eps$ 
divergence of $V$ related to the cusp anomalous dimension via \eq{cusp-def}.
For the computation of the MI we use the {\tt HyperInt} 
package~\cite{Panzer:2014caa}. This code allows to automatically evaluate 
linearly reducible convergent Feynman integrals in terms of multiple 
polylogarithms. In our case the latter reduces to transcendental numbers.%
\footnote{The maximal transcendental weight appearing in $\Gamma_{\rm cusp}$ 
can be roughly estimated as follows:
In ${\mathcal N}=4$ sYM the four-loop $\Gamma_{\rm cusp}(\phi)$ is believed to have uniform 
transcendental weight seven, equal to the maximum degree of divergence (two 
times the loop number) minus one, because it is related to the $1/\eps$ 
coefficient in $W$ ($\eps$ has weight minus one).
Assuming this to be the maximum weight for $\Gamma_{\rm cusp}(\phi)$ in QCD and 
subtracting one in the limit $\phi \to 0$ and one for the $n_f$ piece given the 
experience at lower loops~\cite{Grozin:2015kna}, we arrive at a maximum 
transcendental weight of five.}

In order to provide finite Feynman parameter integrals as an input to {\tt 
HyperInt} we first switch to a MI basis without divergent subintegrals (except for
trivial bubble insertions that can be integrated out.)
In practice this is done by inserting a sufficient number of dots on the 
(off-shell) heavy quark lines of the MI, i.e. by increasing the power of the 
linear propagators by an integer amount. The resulting new basis of MI is then 
related to the old one by IBP reduction.
Next, we Fourier transform to position space and directly integrate out the 
simple bubble and HQET self-energy subintegrals by hand.
This effectively produces integrals with non-integer $\eps$-dependent 
propagator powers, but less than four loops. 
In order to avoid generating new divergent subintegrals in this process, it may 
be necessary to increase the powers also of some of the involved bubble 
propagators beforehand.
The resulting integrals all have at most one overall UV divergence ($\propto 
1/\eps$).

Without loss of generality we can fix the position of the left-most vertex on 
the Wilson line to the coordinate origin ($x_1=0$) and parametrize the 
following vertices from left to right along the Wilson line by $\rho \, x_2$, 
$\rho (x_2 +x_3)$, etc., where $0 \le x_i \le 1$, $\sum_i x_i = 1$. The 
parameter $\rho$ has the dimension of a length and $0 \le \rho \le \infty$.
We can now write down a Feynman parameter representation for the integral over 
the positions of the non-Wilson-line vertices as a function of the $x_i$ and 
$\rho$.
In particular its dependence on the only dimensionful parameter, $\rho$, can be 
easily deduced from dimensional power counting.
The position space representation of a HQET Wilson line propagator between the 
vertices $l$ and $m$ of arbitrary power $a$ reads
\begin{align}
 \int \frac{\df q}{2 \pi} \; \frac{e^{-i q \rho \, x_m}}{ (-2 q + \delta)^a} 
= \left(\frac{i}{2} \right)^{\!\!a} \frac{1}{\Gamma(a)} \, \rho^{a-1} \, 
x_m^{a-1} 
\, e^{-i x_m \rho\, \delta / 2} \,.
\label{eq:HQETProp}
\end{align}
Note that because of the causal $-i0$ prescription $\delta$ can be considered 
to have an infinitesimally small negative imaginary part.
The remaining overall UV divergence, if present, originates from the 
integration region, where all vertices on the Wilson line are contracted to one 
point, i.e. where $\rho \to 0$, cf.~\rcite{Grozin:2015kna}.
In our parametrization, the product of all heavy quark propagators  
in a diagram together with the $\rho$-dependence from the Feynman parameter 
integral thus yields $\rho^z \exp(-i \rho\, \delta / 2)$. The power $z$ only 
depends on $\eps$ and is fixed by the dimensionality of the MI.
The possibly divergent $\rho$ integral can therefore be carried out easily.
We have thus factored out the UV divergences completely 
and are left with a convergent integral over Feynman parameters and the $x_i$.
Its integrand can now safely be expanded in $\eps$. The individual 
terms in this expansion are then computed with {\tt HyperInt}.

Let us illustrate this procedure for the following non-planar nine-propagator 
MI (number 38 in our list):
\begin{align}
&\text{MI}_{38} 
=\!\raisebox{-4 mm}{\includegraphics[width=4 cm]{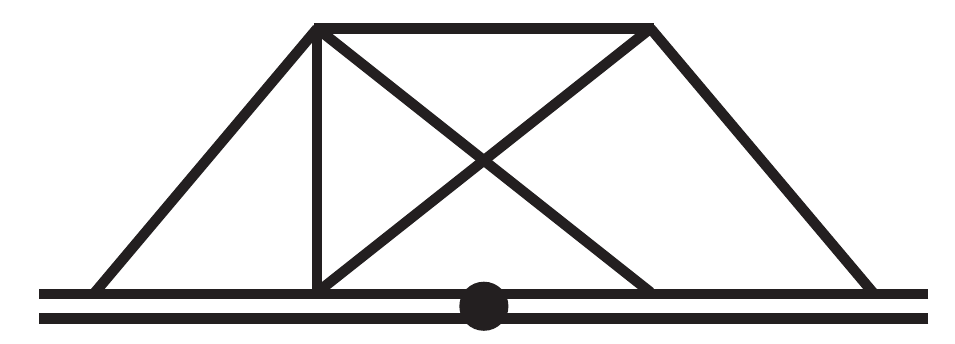}}\!
= G_{1, 2, 1, 1, 0, 1, 1, 0, 1, 1, 0, 1, 0, 0}
= (4\pi)^{2 d} e^{2 \gamma_E  (4-d)}
\!\int_0^\infty \!\! \df \rho \rho^2 \int_0^1 \! \df x_1
\nn\\
&
\times   \int_0^{1-x_1} \!\! \df x_2 \, \bigg[ \frac{1}{16} \, \rho\, x_2  
\, e^{- i \delta \rho /2 } \bigg]
\int \prod_{i=1}^6 \df \alpha_i \bigg[ 
\frac{\pi ^{-2 d}  \Gamma 
(2 d-6)}{4096}  (\alpha_1 \alpha_2 \alpha_3 \alpha_4 
\alpha_5 \alpha_6)^{\frac{d-4}{2}} F^{6-2 d} U^{\frac{3}{2}d-6} \bigg]
,
\label{eq:MI37calc}
\end{align}
where
\begin{align}
F &=  (-\rho^2)\bigg\{
\alpha_1 \alpha_2 (\alpha_3+\alpha_4+\alpha_5+\alpha_6) (1-x_1)^2 + \alpha_6 
\Big(\alpha_1 \alpha_3 x_1^2+\alpha_4 \big[\alpha_2 (1-x_1)^2+\alpha_3 
x_1^2\big]
\nn\\
&\quad +\alpha_5 \big[\alpha_1 x_2^2+\alpha_4 x_2^2 + \alpha_2 
(1-x_1-x_2)^2+\alpha_3 (x_1+x_2)^2\big] +\alpha_2 \alpha_3 \Big)
\nn\\
&\quad +(\alpha_1+\alpha_2) \Big(\alpha_4 \alpha_5 x_2^2+\alpha_3 \big[\alpha_4 
x_1^2+\alpha_5 (x_1+x_2)^2\big]
\Big)
\bigg\} \,,
\label{eq:FPoly}
\\
U &= (\alpha_1+\alpha_2) 
(\alpha_3+\alpha_4+\alpha_5)+(\alpha_1+\alpha_2+\alpha_3 + \alpha_4+ \alpha_5) 
\alpha_6 \,.
\label{eq:UPoly}
\end{align}
We have inserted a dot on the middle heavy quark line in order to render 
the two- and three-loop subintegrals finite.
The first expression in squared brackets in \eq{MI37calc} arises from the 
product of the three Wilson line propagators according to \eq{HQETProp} (for 
the middle one $a=2$).
The second term in squared brackets corresponds to 
the Feynman parameter ($\alpha_i$) representation 
of the integration over the two internal space-time vertices
(the ones not lying on the Wilson line).
Doing the $\rho$ integral yields a factor ($\delta \to \delta - i0$)
\begin{align}
\int_0^\infty \! \df \rho \, \rho^3\, \big(\!-\!\rho^2\big)^{6-2 d}  \, e^{- i 
\delta \rho /2} 
=
 \Gamma (16-4 d) \, \bigg(\frac{\delta}{2} \bigg)^{\!\!4 (d-4)},
\end{align}
which is $1/\eps$ (UV) divergent.
The $\alpha$-integrations in eq. (\ref{eq:MI37calc}) are projective, see e.g. \cite{Smirnov:2006ry}, 
and we use this freedom to choose $\alpha_1 \equiv 1$. With this choice 
the other Feynman parameters are integrated from zero to infinity.
After expansion in $\eps$ we evaluate these convergent integrals together with 
the ones over $x_1$ and $x_2$ using {\tt HyperInt}.
The result is
\begin{align}
\text{MI}_{38}&= \frac{1}{\eps}  
\bigg(
\frac{\pi ^2 \zeta_{3} }{12}+\frac{5 \zeta_{5} }{2}
\bigg)
+5 \zeta_{5} -\frac{17 \zeta_{3}^2}{2}+\frac{\pi ^2 \zeta_{3}}{6}-\frac{169 \pi ^6}{6480}
+\ord{\eps}\,.
\label{eq:MI37res}
\end{align}
Here we have set the IR regulator $\delta = 1$ for convenience. This will 
not affect the final result for the cusp anomalous dimension.
Also the $\ord{\eps}$ term in \eq{MI37res} will not contribute to $\Gamma_{\rm 
cusp}(\phi)$ through $\ord{\phi^4}$, as can be verified by 
inspection of the overall $\eps$-dependent coefficient of MI$_{38}$ in the 
IBP reduced expressions for the vertex function.

With the methods described above we have computed all MI to the relevant order 
in $\eps$.
This, in particular, includes the three eleven-propagator MI corresponding 
to the maximal graphs (with single propagator powers) of the three integral 
topologies in \fig{topos}. For the latter already the leading term in the 
$\eps$ expansion turns out to be sufficient.
We have checked our analytic results numerically with 
{\tt FIESTA}~\cite{Smirnov:2015mct}.
A list of the results for the MI in electronic form can be found in the 
ancillary file of the present paper.

For a number of MI it is straightforward to derive four-fold Mellin--Barnes representations by applying the two-fold representation for the heavy--light vertex~\cite{Davydychev:2001ui}%
\footnote{Note that there is a typo on the right-hand side of eq.~(49) in~\rcite{Davydychev:2001ui}. It should contain an extra factor of two from the Jacobian of the $(t_1,t_2) \to (s,t)$ transformation.}
twice.
They can be transformed to a four-fold series, which can be used to obtain the expansions in $\epsilon$. 
We however find this method less convenient than the procedure described above.

\section{The HQET heavy-quark field anomalous dimension}
\label{sec:HQET}

Let us denote  the heavy quark self energy ($i$ times the sum of all 1PI heavy quark self-energy diagrams) by $\Sigma_h$, and the external heavy quark energy by $\omega$.
In our configuration $\omega= - \delta/2$.
The HQET (off-shell) Ward identity then relates $\Sigma_h$ to the vertex function at zero angle via
\begin{align}
V(0)  &= 1 - \frac{\partial \Sigma_h(\omega)}{\partial \omega}\,.
\label{eq:HQETWardID}
\end{align}
In fact, we have employed this identity already in \eq{lnW}. 
Hence, we can write
\begin{equation}
\log V(0) = - \log Z_h + \ord{\eps^0}\,,
\end{equation}
where $Z_h$ is the ($\MS$) heavy quark wave function renormalization factor and $V(0)$ is expressed in terms of the renormalized coupling $\alpha_s(\mu)$ and gauge parameter $\xi(\mu)$.
We thus have
\begin{align}
\gamma_h \big|_{n_f \Cfour} \equiv   \frac{\df\log Z_h}{\df\log\mu} \Big|_{n_f \Cfour}
= 8 \eps V(0) \big|_{n_f \Cfour} + \ord{\eps} 
\end{align}
for the $n_f \Cfour$ term of the associated anomalous dimension $\gamma_h$, which determines the running of the renormalized HQET heavy quark field $h$ through the renormalization group equation
\begin{align}
\frac{\df}{\df\log\mu}\, h(\mu) = - \frac{\gamma_h}{2}\, h(\mu)\,.
\end{align}
As field renormalization is irrelevant to physical observables $\gamma_h$  can depend on the gauge. The gauge dependence has been explicitly shown at three loops, where the complete result is known~\cite{Melnikov:2000zc,Chetyrkin:2003vi}.
Unlike other parts, the four-loop $n_f \Cfour$ term, however, is gauge invariant for the reason discussed above.
For $\phi=0$ our calculation of the vertex function yields
\begin{equation}
\gamma_h \big|_{n_f \Cfour} =
n_f \Cfour \bigg(\frac{\alpha_s}{\pi} \bigg)^{\!\!4}
\biggl(\!\!
- \frac{5}{4}  \zeta_5
+ \frac{2}{3} \pi^2 \zeta_3
 + \zeta_3
- \frac{2}{3} \pi^2
\biggr)\,.
\end{equation}
We have also calculated the $n_f \mathcal{C}_{F,4}$ term of $\Sigma_h$ from HQET heavy quark self energy diagrams (without cusp) and checked explicitly that \eq{HQETWardID} is fulfilled.

The $\MS$ renormalized QCD heavy-quark field $Q$ is related to the $\MS$ renormalized HQET field $h$ by the matching relation~\cite{Grozin:2010wa}
\begin{align}
Q(\mu) &= z(\mu)^{1/2} h(\mu) + \ord{\frac{1}{m}},
\\
z(\mu) &= \frac{Z_h \Big(\alpha_s^{(n_f)}(\mu),\xi^{(n_f)}(\mu) \Big) Z_Q^{\text{os}} \Big(g_0^{(n_f+1)},\xi_0^{(n_f+1)} \Big)}%
{Z_Q \Big(\alpha_s^{(n_f+1)}(\mu),\xi^{(n_f+1)}(\mu) \Big) Z_h^{\text{os}} \Big(g_0^{(n_f)},\xi_0^{(n_f)} \Big)}\,.
\end{align}
The $Z_i$ and $Z_i^{\rm os}$ denote the renormalization factors for the field $i=Q,h$ in the $\MS$ and on-shell scheme, respectively.
Bare quantities are labeled by a subscript 0 and $\xi(\mu)$ is the $\MS$ renormalized gauge parameter: $1-\xi_0 = Z_A\big(\alpha_s(\mu),\xi(\mu) \big) \big(1-\xi(\mu) \big)$ with $Z_A$ being the $\MS$ renormalization factor of the gluon field.
The superscripts on the couplings and $\xi$ indicate the number of active flavors in the theory.

If we take all $n_f$ light flavors to be massless, we have $Z_h^{\text{os}} = 1$, because the HQET self-energy diagrams are scaleless in the on-shell limit, i.e. for $\omega \to 0$. 
The four-loop $n_f \Cfour$ term in the anomalous dimension $\gamma_Q$ is known~\cite{Czakon:2004bu}:
\begin{align}
\gamma_Q \big|_{n_f \Cfour} 
&\equiv   \frac{\df\log Z_Q}{\df\log\mu} 
\Big|_{n_f \Cfour} =
n_f \Cfour \bigg(\frac{\alpha_s}{\pi} \bigg)^{\!\!4}\,,
\end{align}
and we can thus determine $Z_Q$.
The on-shell heavy-quark field renormalization constant $Z_Q^{\text{os}}$ is known up to three loops~\cite{Melnikov:2000zc}.
As $z(\mu)$ must be finite in the limit $\epsilon\to0$ we then find
\begin{equation}
Z_Q^{\text{os}} \big|_{n_f \Cfour} = n_f \Cfour \frac{g_0^8 m_{\text{os}}^{-8\epsilon}}{(4\pi)^{2d}}
\left[ - \frac{8}{\epsilon} \left( 5 \zeta_5 - \frac{8}{3} \pi^2 \zeta_3 - 4 \zeta_3 + \frac{8}{3} \pi^2 + 4 \right)
+ \mathcal{O}(\epsilon^0) \right]\,.
\label{HQET:Zos}
\end{equation}
This result can be used as a check of a future four-loop calculation of $Z_Q^{\rm os}$.

\section{Result for the cusp anomalous dimension}
\label{sec:results}

Putting all pieces together we obtain
\begin{align}
\Gamma_{\rm cusp} \big|_{n_f \Cfour} ={}& 
n_f \Cfour \bigg(\frac{\alpha_s}{\pi} \bigg)^{\!\!4} 
\frac{1}{9}
\Bigg[
\phi^2 \bigg(\!\!
- 4 \pi^2 \zeta_3
+ \frac{5}{12} \pi^4
+ \frac{5}{6} \pi^2
\bigg)
\nn\\
&{}
+ \phi^4 \bigg(\!\!
- 4 \zeta_5
- \frac{16}{75} \pi^2 \zeta_3
+ \frac{71}{25} \zeta_3
+ \frac{49}{900} \pi^4
- \frac{157}{900} \pi^2
- \frac{23}{100}
\bigg)
+\ord{\phi^6}
\Bigg]\,  \label{eq:resultcusp4} \\
={}& n_f \Cfour \bigg(\frac{\alpha_s}{\pi} \bigg)^{\!\!4}
\Big[0.150721 \,\phi^2 + 0.00965191 \,\phi^4 +\ord{\phi^6} \Big]
\,, \label{eq:resultcusp4numerical}
\end{align}
while the conjecture from \rcites{Grozin:2014hna,Grozin:2015kna} predicts
\begin{align}
\Gamma_{\rm cusp} \big|_{n_f \Cfour}^\text{conjecture} ={}& 
n_f \Cfour \bigg(\frac{\alpha_s}{\pi} \bigg)^{\!\!4}
\frac{1}{192} \biggl( \phi ^2 + \frac{\phi ^4}{15} +\ord{\phi^6} \biggr)
\bigg( 16 \pi ^4 \log^2 2 - 336 \pi^2 \zeta_3 \log2
\nn\\
&{} - \frac{16}{3} \pi^4 \log2 - 32 \pi^2 \log2
+ \frac{488}{3} \pi^2 \zeta_3 - \frac{5}{3} \pi^6 + \frac{92}{3} \pi^4  - \frac{632}{9} \pi^2
\bigg) \label{eq:conjecturecusp4}
\\
={}& n_f \Cfour \bigg(\frac{\alpha_s}{\pi} \bigg)^{\!\!4}
\Big[0.14801 \,\phi^2 + 0.00986736 \,\phi^4 +\ord{\phi^6}\Big] \label{eq:conjecturecusp4numerical}
\end{align}
with input from the analytic three-loop static potential~\cite{Lee:2016cgz}.

We see from our result, \eq{resultcusp4}, that the relative coefficients of the $\phi^2$ and $\phi^4$ terms 
differ from the conjectured form, \eq{conjecturecusp4}.
Moreover, taking into account the analytic form of the static quark anti-quark potential, 
we see that the latter involves transcendental constants such as $\log2$ that 
do not appear in our four-loop MI.
All of this suggests that the full function $\Gamma_{\rm cusp}(\phi)$ takes a more complicated form,
so that it can reproduce these features.

Nevertheless, it is interesting to compare the numerical size of the contributions.
The exact expression in \eq{resultcusp4numerical} and the (wrong) conjecture in \eq{conjecturecusp4numerical} are numerically remarkably close.

\begin{acknowledgments}
This work was supported in part by the Deutsche Forschungsgemeinschaft through the project ``Infrared and threshold effects in QCD'',
by a GFK fellowship and by the PRISMA cluster of excellence at JGU Mainz.
A.G.'s work has been partially supported by the Russian Ministry of Education and Science.
This research was supported by the Munich Institute for Astro- and Particle Physics (MIAPP) of the DFG cluster of excellence ``Origin and Structure of the Universe".
The authors gratefully acknowledge the computing time granted on the supercomputer Mogon at JGU Mainz.
\end{acknowledgments}

\phantomsection
\addcontentsline{toc}{section}{References}
\bibliographystyle{jhep}
\bibliography{Cusp}

\end{document}